\documentstyle[amssymb,thmsa,graphicx]{article}

\input{tcilatex}
\begin{document}

\begin{center}
{\bf rf SUSCEPTIBILITY OF }$La_{1-x}Sr_{x}MnO_{3}${\bf \ SINGLE CRYSTALS:
MAGNETIC SIGNATURES OF STRUCTURAL CHANGES}
\end{center}

\noindent P.V. PARIMI, H. SRIKANTH, M. BAILLEUL, and S. SRIDHAR, 

\noindent Department of Physics, Northeastern University, Boston, MA 02115.
\vskip.2cm
\noindent R. SURYANARAYANAN, L. PINSARD and A. REVCOLEVSCHI. 

\noindent Laboratoire de Chimie des Solides, UA 446, Universit\v{e}
Paris-Sud, Orsay 91405, France.

\medskip

\noindent {\bf ABSTRACT}
\vskip.2cm
A sensitive tunnel diode oscillator (TDO) operating at $4MHz$ is
used to probe the dynamic response of $La_{1-x}Sr_{x}MnO_{3}$ single
crystals for $x=0.125,0.175,0.28$ and $0.33$ doping. Systematics of the
measured change in reactance $(\Delta X)$ as a function of temperature $%
(30K<T<320K)$ and DC magnetic field $(0<H<6kOe)$ reveal distinct temperature
and field scales associated with the dynamic response of spin. It is notable
that these features are far more striking than the corresponding features in
static measurements. The results are discussed in the context of structural
changes leading to polaron ordering.
\vskip.2cm
\noindent {\bf INTRODUCTION}
\vskip.2cm
The perovskite oxides of the form $\func{Re}_{1-x}A_{x}MnO_{3}$ (where $%
\func{Re}$ is a rare earth such as $La$ and $A$ is a divalent element such
as $Sr$ or $Ca$) have generated considerable interest in recent times
because of the discovery of the colossal magnetoresistance (CMR) effect \cite
{CMR}. The CMR is a direct consequence of an unusual paramagnetic insulator
(PMI) to ferromagnetic metal (FMM) transition driven mainly by the double
exchange mechanism \cite{zener}. However, double exchange alone cannot
describe the complete phase diagram of the manganites and it has been
pointed out that the interplay of strong electron-phonon coupling and double
exchange is required to understand the existence of the high temperature
insulating phase, the CMR effect and its sensitivity to magnetic field \cite
{millis}\cite{roder}.

The deficiency of the double exchange model is the fact that it does not
consider spin-lattice or charge-lattice interactions, namely, Jahn-Teller
interactions and polarons \cite{zhou97}. Experimental results clearly
suggest that lattice contributions are important for a thorough
understanding of manganites. Besides MI transition, charge ordering (CO) is
one of the characteristic phenomena observed in these materials especially
in the low doping regime. CO and stripe correlations of concentrated holes
and spins have attracted much attention in recent times, particularly due to
their possible role in high T$_{c}$ superconductivity\cite
{tran95,salkola96,li99}.

A variety of experiments including structural \cite{kawano96}, transport 
\cite{asamitsu96,anane97} and thermal \cite{uhlenbruck99} measurements have
revealed novel features in the $\func{Re}_{1-x}A_{x}MnO_{3}$ directly
associated with the interplay between structural, electronic and magnetic
properties. Most of the experiments on manganites have been {\em static} and
there have been relatively few experiments which probe the {\em dynamic}
response of these systems. Dynamic experiments are likely to provide
significant information about the collective response of spin and charge to
the oscillating electric and magnetic fields impressed on the materials. In
the present work rf dynamic response of La$_{1-x}$Sr$_{x}$MnO$_{3}$ for
concentrations x=0.125, 0.175,0.28 and 0.33 are reported. We focus on the
interplay of between holes and lattice distortions to understand the
relation between the magnetic and structural properties.

\vskip.2cm
\noindent {\bf EXPERIMENT}
\vskip.2cm
Single crystals of La$_{1-x}$Sr$_{x}$MnO$_{3}$ were grown using an image
furnace technique \cite{revco93}. Samples used in these measurements had
cylindrical disk like shapes with diameter $5mm$ and thickness $2mm$ with
polished surfaces and edges. The rf experiments were performed using a
tunnel diode oscillator (TDO) which has very high sensitivity in measuring
the electro- and magneto-dynamic properties of materials. The crystal is
placed inside a copper coil which forms part of an $LC$-tank circuit driven
by a stable tunnel diode oscillator. The inductive coil with the sample is
mounted at the end of a rigid co-axial cable can be inserted into a
continuous flow Helium cryostat. The temperature of this system can be
regulated between $4.2K$ and $320K$ and an electromagnet is used to apply a
dc magnetic field up to $6kOe$. The resonant frequency ($f_{0})$ is
typically in the range of $2-4MHz$ depending on the geometric
characteristics of the inductive coil and sample dimensions. The quantity
that is measured, the change in frequency $\Delta f=f(T,H)-f_{0})$ as a
function of $T$ and $H,$ is proportional to the change in reactance $\Delta
X $. For magnetic metals, from elementary considerations and applying
Maxwell's equations, it can be shown that:$\Delta X\varpropto \sqrt{\chi },$
where $\chi $ is the differential susceptibility, $dM/dH$ of the material.

\begin{figure}[tbph]
\begin{center}
\includegraphics*[width=.8\textwidth]{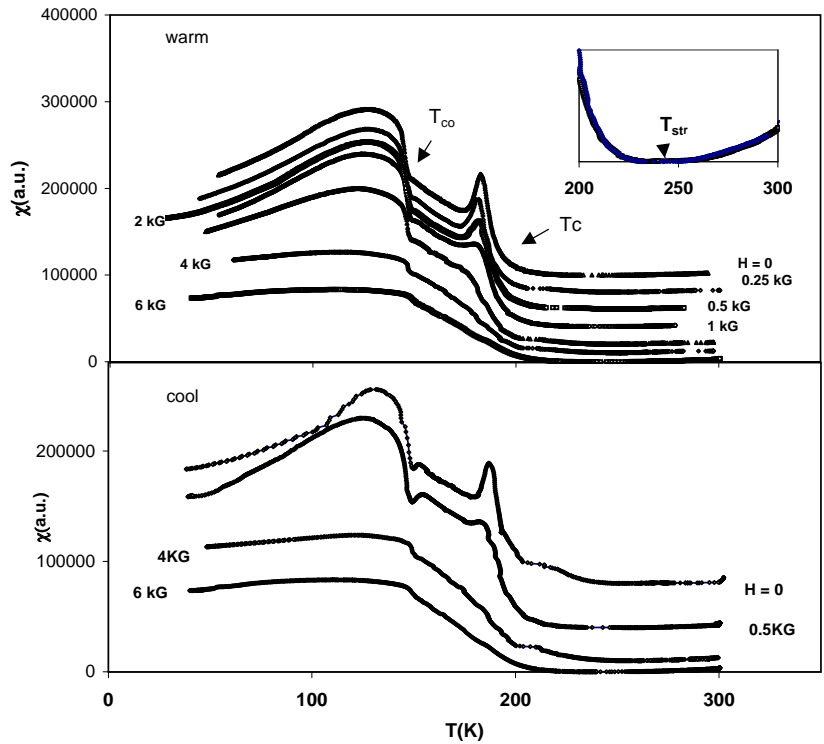}
\end{center}
\par
\caption{Top panel shows differential
susceptibility for x=0.125 doping during warming. Bottom panel shows the
same during cooling. The first order nature of the CO transition can be
seen as a clear difference in the response between warming and cooling.
Inset
shows a magnified view of the structual transition. }
\end{figure}

\noindent {\bf RESULTS}
\vskip.2cm
\noindent {\bf Temperature dependence: La}$_{0.875}${\bf Sr}$_{0.125}${\bf %
MnO}$_{3}$

The high sensitivity of the rf technique enables us to clearly detect a
paramagnetic to ferromagnetic transition at T$_{c}$=180K as well as two
additional transitions at T$_{s}$=270K and T$_{co\text{ }}$=150K, as shown
in Fig. 1. Interestingly, this composition is observed to undergo structural
transitions which are manifested in the change of lattice parameters at
150K, 180K and 270K \cite{pinsard97}. At T$_{s}$ the susceptibility shows a
dip which is due to a structural phase transition from orthorhombic (pseudo
cubic) to a cooperative Jahn-Teller distorted phase at lower temperature. In
the presence of a magnetic field two characteristic changes are observed to
take place at T$_{c}$. First, the peak disappears and secondly, the
transition is broadened. In the absence of dc magnetic field the
susceptibility is zero above $T_{c}$ and raises rapidly at T$_{c}$. In the
presence of dc field the susceptibility is finite at all temperatures and
increases with applied field. Therefore, the sharp transition at T$_{c}$
becomes broadened when field is applied. The hump observed at T$_{co\text{ }%
} $is very clear and strong unlike the CO transition observed in resistivity
and magnetization measurements\cite{uhlenbruck99}. This fact emphasizes the
importance of high frequency measurements to detect CO transitions. The hump
at T$_{co}$ is caused by a magnetic transition accompanied by a change in
structure\cite{uhlenbruck99}\cite{pinsard97,kawano96}.We also observed
hysteretic behavior in the susceptibility around T$_{co}$ with decreasing
and increasing temperature which indicates the first order nature of this
transition. As can be seen from the Fig. 1 the rf reactance shows a dip at T$%
_{co}$ during cooling which is not observed while warming.

The hump associated with T$_{co}$ appears to be a purely ac response of the
charge ordering as the dc response\cite{uhlenbruck99} does not show any
hump. It is worth mentioning that the CO observed in Nd$_{0.45}$Ca$_{0.55}$%
MnO$_{3}$ at 260K also shows a hump in the ac susceptibility measurement\cite
{mukhin98}. The reason for the hump only in ac measurement is that in ac
measurement the differential susceptibility is measured. The reversible
response of the ferromagnetic domains to the rf field just below T$_{c}$
gives rise to an increase in the differential susceptibility. With further
decrease in T the onset of saturation magnetization locks the individual
domain and hence the $\chi (T)$ starts decreasing.

The key to understanding the contribution of the structural transitions to
the electronic and magnetic properties lies in the Mn-O interionic distance
of the octahedra. The interionic distances $m(T),s(T)$, and $l(T),$ which
are along $a,c$ and $b$ axes, respectively, are calculated from the
representation $m^{2}=0.031(a^{2}+b^{2}+c^{2})$, $s^{2}=0.125c^{2}-m^{2}$
and $l^{2}=a^{2}s^{2}/(16s^{2}-a^{2})$. In these calculations the rotation
of the octahedra with respect to the axes is neglected. Fig 2 shows the
temperature dependence of these parameters. As can be seen from the Fig. $%
m(T)$ is constant over the entire temperature range, while $s(T)$ and $l(T)$
show clear anomalies at 140K and 270K. These results imply that for
temperatures below 140K or above 270K there is no contribution of the
rhombic J-T Q2 mode to the formation of crystal lattice. The turning on of
the Q2 mode as the sample is warmed above 140K results in structural phase
transition from low temperature. The response of a ferromagnet in a magnetic
field is also important to describe the first order transitions observed at T%
$_{str}$ and T$_{co}$. Below T$_{str}$ the system shows a spontaneous
cooperative JT distorted phase\cite{uhlenbruck99}. The strong dependence of $%
\chi (T)$ on magnetic field suggests magnetoelastic coupling for CO besides
coulomb repulsion.

\begin{figure}[tbph]
\begin{center}
\includegraphics*[width=.8\textwidth]{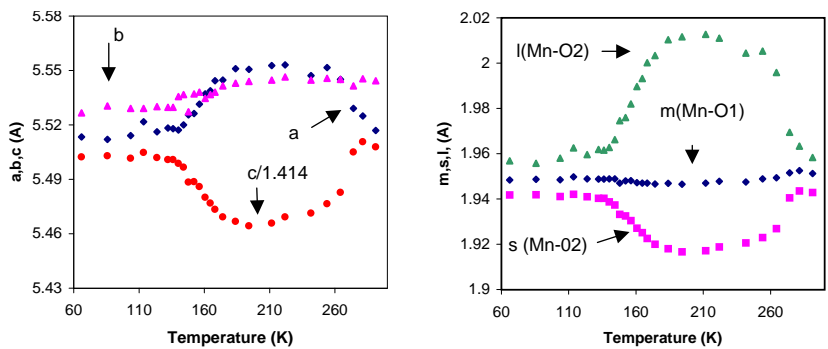}
\end{center}
\caption{The change in the lattice
parameters is shown on the left panel. Right panel shows the Mn-O
interionic
distances, m(T), s(T) and l(T) which are along a,c and b axes. }
\end{figure}

An isolated hole in $LaMnO_{3}$ can be considered a small polaron which is
given by a localized hole in the $3d_{x^{2}-y^{2}}$ orbital surrounded by
inverse Jahn-Teller distortion. The polaron phase is an ordered arrangement
of Mn$^{3+}$ and Mn$^{4+}$ ions for which one of the two alternating atomic
layers in the (001) plane contains both Mn$^{3+}$ ions, as in pure LaMnO3,
while the other layer contains both Mn$^{3+}$ and Mn$^{4+\text{ }}$ions,
i.e. holes. The local distortion is because the hole site Mn$^{4+}$ is JT
inactive whence the electron-phonon energy is lowered by restoring higher
symmetry around the hole\cite{yamada96}. In this picture, at high
temperatures the La$_{1-x}$Sr$_{x}$MnO$_{3}$ may be viewed as a polaron
liquid which will eventually transform into polaron lattice as the
temperature is lowered. We, therefore, identify $T_{co}$ as the onset point
for polaron lattice formation, where holes start to freeze on lattice
points. From this point of view La$_{1-x}$Sr$_{x}$MnO$_{3}$ is considered to
undergo successive transitions from polaron liquid (insulator) to Fermi
liquid (metal) to polaron lattice (insulator).

\begin{figure}[tbph]
\begin{center}
\includegraphics*[width=.8\textwidth]{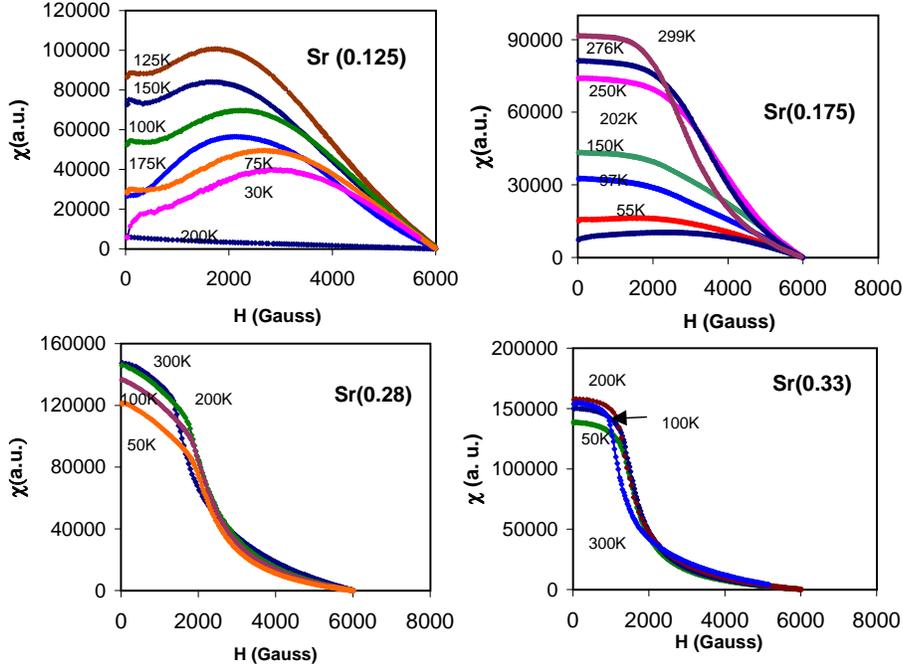}
\end{center}
\caption{(Left) Field dependence of
differential susceptibility for x=0.125, 0.175, 0.28 and 0.33 doping. }
\end{figure}

\noindent \noindent {\bf Field Dependence: La}$_{1-x}${\bf Sr}$_{x}${\bf MnO}%
$_{3}$

The field dependence of the differential susceptibility, $\frac{dM}{dh_{ac}}%
\mid _{H_{dc}},$ at various temperatures below 300K is shown in Fig. 3. As
can be seen from the figure for all the doping levels of Sr the $\chi (H)$
response shows an overall decrease with the increase in magnetic field. The
magnetization M(H) shows a monotonic increase with field with an eventual
saturation at high fields, for all the compositions studied. Therefore, the
decrease in the differential susceptibility is not surprising. There are,
however, many subtle changes in the $\chi (H)$ response at low magnetic
fields, H \TEXTsymbol{<} 2000G. In the case of x=0.125 composition $\chi (H)$
initially increases, reaches a maximum and starts decreasing forming a peak.
For the remaining three compositions the peak is not prominent as can be
seen from the figure. The behavior of $\chi (H)$ can be understood by a
simple picture of domain response to weak and strong magnetic fields. When a
weak field is applied the magnetization process is reversible. In the
presence of strong fields the domains are locked and tend to form a single
domain. Therefore, the response of the domains to the ac field at low dc
fields is greater than that at high dc fields, thus contributing to the
initial increase.
\vskip.2cm
\noindent {\bf CONCLUSIONS}
\vskip.2cm
Dynamic rf susceptibility of La$_{1-x}$Sr$_{x}$MnO$_{3}$ revealed several
magnetic signatures in both temperature and field dependent measurements.
These magnetic signatures have direct correlation with structural changes in
terms of Mn-O interionic distances of the octahedra, at the corresponding
temperatures. Field dependent differential susceptibility is found to
decrease monotonically with field with a rich structure at low fields.
\vskip.2cm
\noindent {\bf ACKNOWLEDGMENT}
\vskip.2cm
This work was supported by a US-NSF- 9711910 and NSF-CNRS grant
NSF-INT-9726801.

\end{document}